\documentclass[11pt]{article}
\usepackage{amssymb}
\usepackage{amsmath}

 \usepackage{ifthen}

\ifx\pdfoutput\undefined 
\usepackage[ps2pdf,                
bookmarks=true,
bookmarksnumbered=true,
hypertexnames=false,
breaklinks=true,
linkbordercolor={0 0 1},
pdfborder={0 0 112.0}]{hyperref}
\else 
\usepackage[pdftex,                
bookmarks=true,
bookmarksnumbered=true,
hypertexnames=false,
breaklinks=true,
linkbordercolor={0 0 1}]{hyperref} 
\fi 

\allowdisplaybreaks[3]

\makeatletter
\@addtoreset{equation}{section}
\makeatother

\def\dalemb#1#2{{\vbox{\hrule height .#2pt
        \hbox{\vrule width.#2pt height#1pt \kern#1pt
                \vrule width.#2pt}
        \hrule height.#2pt}}}

\def\cA{{\cal A}}

\def\0{{\sst{(0)}}}
\def\1{{\sst{(1)}}}
\def\2{{\sst{(2)}}}
\def\3{{\sst{(3)}}}
\def\4{{\sst{(4)}}}
\def\5{{\sst{(5)}}}
\def\6{{\sst{(6)}}}
\def\7{{\sst{(7)}}}
\def\8{{\sst{(8)}}}
\def\n{{\sst{(n)}}}

\def\hp{ \frac{1}{2}}

\let\a=\alpha \let\b=\beta \let\g=\gamma \let\d=\delta \let\e=\epsilon
    \let\k=\kappa
\let\l=\lambda \let\m=\mu \let\n=\nu  \let\r=\rho
\let\s=\sigma \let\t=\tau    
\let\w=\omega    \let\L=\Lambda
    
 \let\W=\Omega   \let\G=\Gamma

 \def\bd{\begin{document}} \def\ed{\end{document}}
\def\ds{\documentstyle} \let\fr=\frac \let\bl=\bigl \let\br=\bigr
\let\Br=\Bigr \let\Bl=\Bigl
\let\bm=\bibitem
\let\na=\nabla
\let\pa=\partial \let\ov=\overline
\newcommand{\be}{\begin{equation}}
\newcommand{\ee}{\end{equation}}
\def\ba{\begin{array}}
\def\ea{\end{array}}
\def\ft#1#2{{\textstyle{{\scriptstyle #1}\over {\scriptstyle #2}}}}
\def\fft#1#2{{#1 \over #2}}
\def\del{\partial}
\def\sst#1{{\scriptscriptstyle #1}}
 \def\oneone{\rlap 1\mkern4mu{\rm l}}
\def\ie{{\it i.e.\ }}
\def\via{{\it via}}
\def\semi{{\ltimes}}
\def\str{{\rm str}}
\def\Dm{{{D_{\sst{max}}}}}
\def\vac{ \left | 0 \right \rangle }
\def\kvac{ \left | k \right \rangle }

\def\sp{\; \; \;}

\def\bol{ \left | B (p^+) \right \rangle}
\def\bo1{ \left | B^0 (p^+) \right \rangle}

\def\bolt{ \left | B (p^+) \right \rangle_{\t}}

\def\boxl{ \left | B (x^-) \right \rangle}

\def\<{ \langle }
\def\>{ \rangle }

\def\vf{\varphi}

\def\ls{{(l,0)}}
\def\lv{{(l,\pm1)}}
\def\lt{{(l,\pm2)}}

\def\lse#1{{(l_{#1},0)}}
\def\lve#1{{(l_{#1},\pm1)}}
\def\lte#1{{(l_{#1},\pm2)}}

\def\lsg#1{{5(l_{#1},0)}}
\def\lvg#1{{5(l_{#1},\pm1)}}
\def\ltg#1{{5(l_{#1},\pm2)}}

\def\lsi#1{{5{(#1,0)}}}
\def\lvi#1{{5{(#1,\pm1)}}}
\def\lti#1{{5{(#1,\pm2)}}}

\def\lsr#1{{1{(#1,0)}}}
\def\lvr#1{{1{(#1,\pm1)}}}
\def\ltr#1{{1{(#1,\pm2)}}}

\def\cD{{\cal D}}
\def\cE{{\cal E}}
\def\cF{{\cal F}}
\def\cG{{\cal G}}
\def\cH{{\cal H}}
\def\cK{{\cal K}}
\def\cO{{\cal O}}
\def\cP{{\cal P}}
\def\cQ{{\cal Q}}
\def\cR{{\cal R}}
\def\cS{{\cal S}}
\def\cT{{\cal T}}
\def\cU{{\cal U}}
\def\cV{{\cal V}}
\def\cW{{\cal W}}

\newcommand{\nono}{\nonumber}
\newcommand{\dtilde}[1]{\tilde{\tilde{#1}}}
\newcommand{\hatb}[1]{\hat{\ov{#1}}}
\newcommand{\hatt}[1]{\hat{\tilde{#1}}}
\newcommand{\emnr}{{e_\m}^{\n\r}}
\newcommand{\sub}[1]{\phantom{}_{(#1)}\phantom{}}
\newcommand{\rt}{\tilde{\r}}

\newcommand{\comment}[1]{}

\def\hna{\hat{\na}}

\newcommand{\hsp}{\hspace{0.5cm}}

\newcommand{\ho}[1]{$\, ^{#1}$}
\newcommand{\hoch}[1]{$\, ^{#1}$}
\newcommand{\bea}{\begin{eqnarray}}
\newcommand{\eea}{\end{eqnarray}}
\newcommand{\ra}{\rightarrow}
\newcommand{\lra}{\longrightarrow}
\newcommand{\Lra}{\Leftrightarrow}
\newcommand{\ap}{\alpha^\prime}
\newcommand{\bp}{\tilde \beta^\prime}
\newcommand{\tr}{{\rm tr} }
\newcommand{\Tr}{{\rm Tr} }
\newcommand{\NP}{Nucl. Phys. }

\newcommand{\ams}{{\it Institute for Theoretical Physics,
University of Amsterdam, \\
Valckenierstraat 65, 1018XE Amsterdam, The Netherlands} \\
{\tt I.R.G.Kanitscheider, K.Skenderis@uva.nl}}
\newcommand{\auth}{{\large Ingmar Kanitscheider and Kostas Skenderis}}

\begin{document}
\begin{flushright}
\hfill{ITFA-2009-01}\\
\hfill{NSF-KITP-2009-02}
\end{flushright}

\vspace{15pt}

\begin{center}

{\Large \bf Universal hydrodynamics of non-conformal branes}

\vspace{20pt}

\auth

\vspace{15pt}

\vspace{8pt}

{\ams}

\vspace{15pt}

\underline{ABSTRACT}
\end{center}
We examine the hydrodynamic limit of non-conformal branes using the recently developed precise holographic dictionary. We first streamline the discussion of holography for backgrounds that asymptote locally to
non-conformal brane solutions by showing that all such
solutions can be obtained from higher dimensional asymptotically locally AdS solutions by suitable dimensional reduction and continuation in the dimension. As a consequence, many holographic results for such backgrounds follow from the corresponding results of the Asymptotically AdS case. In particular,
the hydrodynamics of non-conformal branes is fully determined in terms of conformal hydrodynamics. Using previous results on the latter we predict the form of the non-conformal hydrodynamic stress tensor to second
order in derivatives. Furthermore we show that the ratio between bulk and shear viscosity is fixed by the generalized conformal structure to be $\zeta/\eta = 2(1/(d-1) - c_s^2)$, where $c_s$ is the speed of sound in the fluid.

\pagebreak

\section{Introduction}

The AdS/CFT correspondence provides a powerful tool to study strongly coupled quantum field theories. In particular, it has become possible to explore strongly coupled finite temperature conformal field theories by analyzing asymptotically AdS black hole backgrounds, identifying the Hawking temperature of the black hole with the temperature of the dual field theory.

 Since any interacting field theory locally equilibrates at high enough densities, it is expected that the evolution of long-wavelength fluctuations of strongly coupled field theories is governed by fluid dynamics.
Recently, it was shown that solutions to the long-wavelength fluctuation equations around the boosted black D3 brane geometry can be mapped to solutions to non-linear equations of hydrodynamics in the dual strongly coupled conformal field theory \cite{Bhattacharyya:2008jc}. Using the well-known AdS/CFT dictionary, the fluctuations of the metric were shown to be dual to a fluid configuration which is determined by a conserved hydrodynamic stress tensor. Demanding the bulk fluctuations to be smooth in the interior constrained the transport coefficients of the dual stress tensor.

As the hydrodynamic limit is a limit of long wavelength fluctuations, one proceeds hereby in a derivative expansion of the velocity and temperature field of the fluid. In \cite{Bhattacharyya:2008jc} the hydrodynamical energy-momentum tensor was computed to second derivative order. In \cite{VanRaamsdonk:2008fp, Bhattacharyya:2008xc,Haack:2008cp, Bhattacharyya:2008mz}, this connection was further explored for pure gravity in arbitrary dimension, in \cite{Erdmenger:2008rm, Banerjee:2008th, Haack:2008xx} for Einstein-Maxwell theory in 4+1 dimensions, and in \cite{Bhattacharyya:2008ji} for gravity coupled to a scalar in 4+1 dimensions. All these cases dealt with geometries which are asymptotically AdS.

In this note, we generalize this discussion to non-conformal branes. More precisely we will
consider solutions that asymptote locally to the
near-horizon limit of D$p$-brane ($p \neq 5$) and fundamental string solutions. We will call
these backgrounds {\it asymptotically non-conformal brane} backgrounds.
Recall that the near-horizon limit of the D$p$-branes corresponds to a decoupling limit
\cite{Itzhaki:1998dd} so one expects at least for these cases
a meaningful gauge/gravity duality and indeed one can set up holographic renormalization for such backgrounds  \cite{Wiseman:2008qa, Kanitscheider:2008kd}. This makes possible to generalize the map of bulk gravity equations to boundary hydrodynamic equations. Earlier computations of transport coefficients of non-conformal brane
backgrounds using linear response theory, following the original work \cite{Policastro:2001yc,Policastro:2002} for conformal backgrounds, include \cite{Benincasa:2006ei, Mas:2007ng, Buchel:2007mf,Natsuume:2007ty} (see also \cite{Parnachev:2005hh,Benincasa:2005iv,Gubser:2008sz} for computations of transport
coefficients of other non-conformal backgrounds).

In the near-horizon limit, the supergravity solutions of D$p$-branes ($p \neq 5$) and fundamental strings are conformal to $AdS_{p+2}\times S^{8-p}$ and exhibit a running dilaton. The universal sector of these backgrounds is obtained by dimensionally reducing on the sphere and truncating to $(p+2)$-dimensional gravity coupled to a scalar. The action and solution can be best analyzed in the {\it dual frame} in which the equations of motion admit a linear dilaton $AdS_{p+2}$ solution \cite{Boonstra:1998mp}. Actually there is
a family of $(d+1)$-dimensional bulk actions
parametrized by the positive parameter $\s$ whose equations of motion admit a linear
dilaton $AdS$ solution, namely the
metric is exactly  $AdS_{d+1}$ and the dilaton is given by a power of the radial coordinate,
with the parameter $\s$ determining the power \cite{Kanitscheider:2008kd}.
When expanding solutions which asymptotically  (locally) approach such solutions in a Fefferman-Graham expansion, $\s$ turns out to also correspond to the radial power of the normalizable mode.

The main observation of this paper is that for half-integer $\s$, the $(d+1)$-dimensional action can be obtained by dimensionally reducing $(2\s+1)$-dimensional pure gravity with cosmological constant on a torus. In particular, any $(d+1)$-dimensional solution which is asymptotically locally  linear dilaton $AdS_{d+1}$ in the dual frame can be lifted to an asymptotically locally $AdS_{2 \s +1}$ pure gravity solution. Furthermore, the lower dimensional equations of motion depend smoothly on $\s$, which implies that every solution can be generalized to arbitrary positive $\s$. In our case we use this fact to obtain a black brane solution which solves the equations of motion with arbitrary $\s$.

More generally, we show that all holographic results derived in \cite{Kanitscheider:2008kd}, namely counterterms, 1-point functions and Ward identities, can be obtained from their well-known counterparts \cite{deHaro:2000wj} using this procedure. Note that in \cite{Kanitscheider:2008kd} we already  saw such a relationship between the holographic results for IIA fundamental strings and M2 branes and D4/M5 branes.
In these cases however this was a manifestation of the M-theory unlift, whereas it is unclear what is the underlying reason for the more general relation we uncover here.

Non-conformal branes admit a generalized conformal structure both at weak and at strong coupling
\cite{Jevicki:1998yr,Jevicki:1998qs,Jevicki:1998ub,Kanitscheider:2008kd}. For example, for the
case of D$p$ branes, the low-energy world-volume theory, namely maximally supersymmetric Yang-Mills theory in $(p+1)$ dimensions, is Weyl invariant when coupled to a background
metric provided the coupling constant is promoted to a background field that transforms appropriately \footnote{In a flat background, the generalized conformal structure implies that the theory
is invariant under generalized conformal transformations
which act not only on the fields in the Lagrangian but also on the coupling constant.}.
This invariance leads to a dilatation Ward identity, which is of exactly the same form as the dilatation
Ward identity for RG flows induced by a relevant operator. The main difference
is that in the latter case the theory flows in the UV to a fixed point and the generalized conformal structure
is inherited from the conformal structure of the fixed point but in the non-conformal theories
the running, within the regime of validity of the corresponding descriptions (weak or strong), is due to dimensionality of the coupling constant. For IIA fundamental strings and
D4 branes, which have $\s=3/2,3$, respectively, a new dimension, the M-theory dimension, opens up at strong coupling and the theories indeed flow to a fixed point. In these cases
one can understand the generalized conformal structure as descending from the conformal symmetry of the corresponding M-theory system. The results of this papers
connect the generalized conformal structure of all cases to a conformal structure of a theory in higher
(albeit non-integral!) dimensions.

Applied to hydrodynamics we can draw the following conclusion. According to \cite{Bhattacharyya:2008jc} and their generalizations, every smooth metric fluctuations around the black brane solution in $(2\s+1)$-dimensional pure gravity can be mapped to a solution of conformal hydrodynamics with specific transport coefficients in $(2\s)$ dimensions. In a similar manner, metric and scalar fluctuations around the non-conformal black brane with given $\s$ in $(d+1)$ bulk dimensions will be dual to a solution to non-conformal hydrodynamics in $d$ dimensions. Since every solution of the non-conformal gravity/scalar system can after continuation in $\s$ be uplifted to a solution of the higher dimensional pure gravity system, we conclude that the non-conformal hydrodynamics in $d$ dimensions can be obtained by dimensional reduction of conformal hydrodynamics in $(2\s)$ dimensions and continuation in $\s$.

An immediate consequence of the fact that the non-conformal hydrodynamic stress tensor can be obtained by dimensional reduction from the conformal stress tensor is that the ratio between bulk and shear viscosity $\zeta/\eta$ is fixed. A different ratio in the non-conformal fluid would uplift to a non-vanishing bulk viscosity in the conformal fluid, which is forbidden by conformal symmetry. A related argument was 
presented in \cite{Gubser:2008sz}.
Hence the ratio of bulk and shear viscosity in the non-conformal fluid is dictated by the generalized conformal structure. Furthermore the ratio we find, and which was found earlier for D$p$-branes in \cite{Benincasa:2006ei, Mas:2007ng, Buchel:2007mf}, saturates the bound proposed in \cite{Buchel:2007mf}, $\zeta/\eta \geq 2(1/(d-1) - c_s^2)$, where $c_s$ is the speed of sound in the fluid. This bound was proposed to be universal for strongly coupled gauge theory plasmas,
similar to the KSS conjecture \cite{KSSbound}, $\eta/s \geq 1/4 \pi$, for the the ratio between shear viscosity and entropy density. However, there is an important qualitative difference between the two cases. In the latter case, the ratio that saturates the bound, $\eta/s = 1/4 \pi$, is obtained by requiring smoothness of the bulk solution in the interior, so it has a dynamics origin, whereas in the latter case, the ratio that saturates the bound, $\zeta/\eta = 2(1/(d-1) - c_s^2)$, follows from the generalized conformal structure
so it is of kinematical origin.

In \cite{Bhattacharyya:2008jc} and generalizations, the transport coefficients of the conformal hydrodynamic stress tensor were computed by demanding smoothness in the interior. We use the map between conformal hydrodynamics and hydrodynamics of generalized non-conformal branes to predict the form of the stress tensor to second derivative order. In the sequel, we confirm the first order form by an independent bulk calculation. However, instead of using the framework of \cite{Bhattacharyya:2008jc}, in which the bulk equations were analyzed in Eddington-Finkelstein coordinates, we use the method of \cite{Gupta:2008th} where the black D3 brane fluctuations were analyzed in Fefferman-Graham coordinates. This is advantageous since it allows for a Lorentz covariant expansion, the constraint equations become trivial and reading off the stress tensor from the bulk metric is completely straightforward. Only to fix the coefficients of the dual stress tensor one switches to Eddington-Finkelstein coordinates, which is necessary to ensure the absence of singularities at the horizon of the black brane. We first generalize the discussion of \cite{Gupta:2008th} to arbitrary dimension $(2\s)$ and compactify to obtain the non-conformal case for general $\s$.

This paper is organized as follows. We begin in section \ref{two} by showing that the gravity/scalar system
relevant for the non-conformal branes
can be obtained by dimensional reduction from pure gravity in $(2\s+1)$ dimensions on a torus and continuation in $\s$,
and discussing the implications of this for holography.
In section \ref{three}, we apply this reasoning to hydrodynamics and conclude that
the hydrodynamics of non-conformal branes  can be obtained by dimensional reduction from conformal hydrodynamics. We predict the form of the non-conformal hydrodynamic energy-momentum tensor up to second derivative order, confirm that the KSS bound \cite{KSSbound} is saturated and comment on Buchel's bound \cite{Buchel:2007mf} for the ratio between shear and bulk viscosity.
Sections \ref{five} to \ref{seven} are devoted to explicitly checking the first order coefficients of the non-conformal energy-momentum tensor by a bulk calculation for the case of flat boundary metric.
In section \ref{five}, we set up the conformal black brane solution and its non-conformal generalization.
In section \ref{six}, we bring the black brane solution to Fefferman-Graham coordinates and calculate its first order correction in the derivative expansion. From that we extract the energy-momentum tensor to first derivative order.
In section \ref{seven}, we transform the first order solution to Eddington-Finkelstein coordinates to examine the singularity at the horizon of the unperturbed black brane. Analogously to \cite{Gupta:2008th}, we find that the solution is only smooth in Eddington-Finkelstein coordinates if the shear viscosity given in the stress tensor saturates the KSS bound. Throughout sections \ref{five} to \ref{seven}, we often make use of the fact that at any point in the calculation we can obtain the non-conformal case by dimensional reduction and continuation from the conformal case.
Finally we end with conclusions and prospects for future research.
In appendix \ref{appA} we discuss the Fefferman-Graham expansion beyond the normalizable mode order
and the dependence of the coefficients on the vev of the energy-momentum tensor.

\section{Lower dimensional field equations} \label{two}
\label{lowdfe}


In the near-horizon limit, the supergravity solutions of D$p$-branes and fundamental strings are conformal to $AdS_{p+2}\times S^{8-p}$ and exhibit a running dilaton. There is a Weyl transformation to the {\it dual frame}, in which the metric is exactly $AdS_{p+2}\times S^{8-p}$ \cite{Boonstra:1998mp}. Reducing the action in the dual frame on the sphere yields \cite{Boonstra:1998mp, Kanitscheider:2008kd}
\be
  \label{actionorigform}
  S = -L \int d^{d+1}x \sqrt{g} e^{\g \phi} [R + \b (\pa \phi)^2 + C],
\ee
where the constants $(L,\b,\g,C)$ depend on the case of interest and are given in \cite{Kanitscheider:2008kd}. The equations of motions admit a linear dilaton $AdS_{d+1}$ solution
\bea
\label{origAdSsolution}
   ds^2 &=& \frac{d\r^2}{4\r^2} + \frac{dz_idz^i}{\r}; \nono \\
   e^\phi &=& \r^\a,
\eea
where $i=1,\ldots,d$, provided that $\a$ and $C$ satisfy
\be
  \a = -\frac{\g}{2(\g^2-\b)}, \qquad C = \frac{(d(\g^2-\b)+\g^2)(d(\g^2-\b +\b)}{(\g^2-\b)^2}.
\ee
Taking $\a$ instead of $\b$ as fundamental parameter we obtain the simpler form
\be
   \b = \g^2(1+\frac{1}{2\a\g}), \qquad C = (d-2\a\g)(d-2\a\g-1).
\ee
After rescaling the scalar $\phi \rightarrow \phi/\g$ the action \eqref{actionorigform} takes the form
\be
\label{lowdaction}
  S = -L \int d^{d+1} x \sqrt{-\hat{G}} e^\phi [R + (1+\frac{1}{2\a\g})(\pa \phi)^2 +(d-2\a\g)(d-2\a\g-1)],
\ee
with solution \eqref{origAdSsolution} becoming
\bea
   \label{AdSsolution}
   ds^2 &\equiv& \hat{G}_{MN} dx^M dx^N = \frac{d\r^2}{4\r^2} + \frac{dz_idz^i}{\r}; \nono \\
   e^\phi &=& \r^{\a\g}.
\eea
We observe that in any dimension $d$ we have a family of actions of the form \eqref{lowdaction} which depend on the parameter $\a\g$ and whose equation of motions each admit a linear dilaton $AdS_{d+1}$ solution of the form \eqref{AdSsolution}. The limit $\a\g \rightarrow 0$\footnote{To be precise this is only a well-defined limit of the action if we first rescale the scalar $\phi \rightarrow \a\g \phi$ before taking $\a\g \rightarrow 0$.} and the choice $\a\g=-(d-4)^2/2(6-d)$ correspond to Einstein frame pure gravity with cosmological constant and decoupled D$p$-branes with $d=p+1$, respectively.
For further reference let us also comment on the slightly non-standard dimensions of solution \eqref{AdSsolution} and action \eqref{lowdaction}. The $AdS$ radius in \eqref{AdSsolution} is absorbed in Newton's constant in the prefactor $L$ of the action \eqref{lowdaction} and furthermore we have length dimensions:
\bea
   \ [\r]&=& 2, \qquad [z^i] = 1, \qquad [ds^2] = -2, \\
   \ [R] &=& 0, \qquad [\sqrt{-\hat{G}}] = -d-2, \nono \\
   \ [\int d^{d+1} x] &=& [\int d\r \, d^d z] = d+2, \nono\\
   \ [ e^\phi ] &=& -[L] = 2\a\g. \nono
\eea

Extracting precise boundary theory data from the asymptotics of decoupled non-conformal brane backgrounds requires using holographic renormalization \cite{Skenderis:2002wp}, which was developed for D$p$-brane backgrounds in \cite{Wiseman:2008qa, Kanitscheider:2008kd}. In \cite{Kanitscheider:2008kd} it was noted that the framework of holographic renormalization could be generalized to arbitrary values of $\a\g$. With the ansatz for metric and scalar
\bea
\label{metrscalans}
  ds^2 &=& \frac{d\r^2}{4\r^2} + \frac{g_{ij}(z,\r)dz^idz^j}{\r}, \\
  \phi(z,\rho) &=& \a\g \log \r + \k(z,\r), \nono
\eea
the field equations following from \eqref{lowdaction} become
\bea
   &&\hspace{-0.5cm}    -\frac{1}{4} \Tr(g^{-1}g')^2 + \hp \Tr g^{-1} g''
+ \k'' - \frac{1}{2\a\g}(\k')^2 = 0, \label{tra-eq} \\
&& \hspace{-0.5cm}    -\hp\na^i {g'}_{ij} + \hp \pa_j
(\Tr g^{-1} g') -\frac{1}{2\a\g}\pa_j \k \k' +
\pa_j \k' - \hp {g'_j}^k \pa_k \k = 0, \label{div-equ} \\
&& \hspace{-0.5cm}    \left[-Ric(g) -2(\s-1)g' - \Tr(g^{-1}g')g + \r(2g''
- 2g'g^{-1}g'+ \Tr(g^{-1}g')g')\right]_{ij}    \nono \\
&&    + \na_i \pa_j \k -
    \frac{1}{2\a\g}\pa_i \k
\pa_j \k -2(g_{ij} - \r g'_{ij}) \k' = 0,   \label{Ricasympt} \\
&&\hspace{-0.5cm}
4\r(\k'' + (\k')^2) +2(d-4\s+2)\k' + \na^2 \k + (\pa \k)^2
    + 2\Tr(g^{-1} g')(\a\g + \r\k') = 0 \label{scalasympt},
\eea
where differentiation with respect to $\r$ is denoted with a prime, $\na_i$ is the covariant derivative constructed from the metric $g$ and $\s \equiv d/2-\a\g$. Since the field equations are polynomials in $\r$ we can conclude that $g(z,\r)$ and $\k(z,\r)$ are regular functions of $\r$ and expand them in powers of $\r$. Inserting these expressions in the equations of motion yields algebraic expressions for the subleading terms $g\sub{2n>0}$ and $\k\sub{2n>0}$ in terms of the sources $g\sub{0}$ and $\k\sub{0}$, until order $n=\s \equiv d/2 - \a\g$, at which only the divergence of $g\sub{2\s}$ and the combination $\Tr g\sub{2\s} + 2 \k\sub{2\s}$ is determined. Furthermore, if $\s$ is integer, we have to introduce logarithmic terms at order $\s$ to fulfill the equations of motion:
\bea
\label{metrscalexp}
  g(z,\rho) &=& g\sub{0}(z) + \r g\sub{2}(z) + \ldots + \r^\s (g\sub{2\s}(z) + h\sub{2\s}(z) \log \r) + \ldots,  \\
  \k (z,\rho) &=& \k\sub{0}(z) + \r \k\sub{2}(z) + \ldots + \r^\s(\k\sub{2\s}(z) + \tilde{\k}\sub{2\s}(z) \log \r) + \ldots. \nono
\eea
 Compared to pure gravity AdS, where $\s=d/2$, the power of the non-local term is shifted by $-\a\g$; this corresponds precisely to the additional factor $e^\phi$ in the action \eqref{lowdaction} ensuring that all counterterms can still be defined as local functionals of the sources and that the non-local terms $g\sub{2\s}$ and $\k\sub{2\s}$ contribute only to the finite part of the regularized action. The presence of the logarithmic terms in \eqref{metrscalexp} appearing for $\s$ integer corresponds precisely to the presence of an
 anomaly \cite{Henningson:1998gx} in the generalized conformal Ward identity of the dual theory:
\bea
\label{Wardid}
   \<T^i_i\> + 2\a\g \<\cO_\phi\> = \cA.
\eea

We will now present a new and much simpler derivation of the holographic results for the non-conformal branes.
Let us first consider the case of half integer $\s>d/2$. In this case
the action \eqref{lowdaction} can be obtained by reducing $(2\s+1)$-dimensional gravity with cosmological constant $\L = -\s(2\s-1)$ on a $(2\s-d)$-dimensional torus with the reduction ansatz
\be
\label{redans}
   ds^2 = ds^2_{(d+1)}(\r,z) + e^{\frac{2\phi(\r,z)}{2\s-d}}dy_a dy^a,
\ee
where $a=1,\ldots ,(2\s-d)$ runs over the torus directions. The Ricci scalar and the action reduce as
\bea
\label{Ricactred}
   R_{2\s+1} &=& R_{d+1} - 2 \na^2 \phi - \frac{2\s-d+1}{2\s-d} (\pa \phi)^2, \\
   S &=& -L_{AdS} \int d^{2\s+1}x \sqrt{-g_{2\s+1}} (R_{2\s+1} + 2\s(2\s-1)) \nono \\
     &=& -L_{AdS} (2 \pi R_y)^{2\s-d}\int d^{d+1} x \sqrt {-g_{d+1}} e^\phi(R_{d+1} + \frac{2\s -d -1}{2\s-d} (\pa \phi)^2 +  2\s(2\s-1)), \nono
\eea
where the (in our conventions) dimensionless prefactor $L_{AdS}$ of the $(2\s+1)$-dimensional pure gravity action is given by
\be
   L_{AdS} = \frac{l_{AdS}^{2\s-1}}{16 \pi G_{2\s+1}},
\ee
with $l_{AdS}$ the radius of the $(2\s+1)$-dimensional AdS space, $G_{2\s+1}$ Newton's constant in $(2\s+1)$ dimensions and $R_y$ the radius of the torus. Given that $\s = d/2-\a\g$ the last line in \eqref{Ricactred} can easily seen to be proportional to \eqref{lowdaction} and thus lead to the same equation of motions. Furthermore one can make the prefactors match by choosing a torus radius $R_y$ so that
\be
\label{llads}
    L = L_{AdS} (2 \pi R_y)^{2\s-d}.
\ee
Thus, since for half-integer $\s>d/2$ the action can be obtained by dimensional reduction, local counterterms
for the action \eqref{lowdaction} can be obtained by reducing the local $AdS_{(2\s+1)}$ counterterms.

Furthermore, the generalized conformal Ward identity \eqref{Wardid} can also be shown to be the dimensional reduction of the conformal Ward identity of $AdS_{(2\s+1)}$. In the conformal case, the vev of the energy-momentum tensor is given by \cite{deHaro:2000wj},
\be
\label{conf_dict}
   \<T_{\m\n}\>_{2\s} = \frac{2}{\sqrt{-g_{(0),2\s}}} \frac{ \d S_{ren}}{\d g_{(0)}^{\m\n}} = 2\s L_{AdS}g_{(2\s) \m\n}  + \ldots,
\ee
where $S_{ren}$ denotes the renormalized on-shell action and the dots denote terms that locally depend
on $g_{(0)\m\n}$. These terms are present when $g_{(0)\m\n}$ is curved and there is a conformal anomaly, i.e.
when $\s$ is an integer. They do not play an important role in the discussion here and so they
will be suppressed. When relating the vev in \eqref{conf_dict} to the vev of the dimensionally reduced theory, we have to account for the additional prefactor $(2\pi R_y)^{2\s-d}$ of the lower-dimensional action in \eqref{Ricactred} which results from the integration over the torus and for the change in the determinant of the metric in the definition of the vev, $\sqrt{g_{(0),d}} = e^{-\k\sub{0}} \sqrt{g_{(0),2\s}}$. One obtains
\bea
   e^{\k\sub{0}} (2\pi R_y)^{2\s-d} \<T_{ij}\>_{2\s} &=& 2\s L e^{\k\sub{0}} g_{(2\s)ij} + \ldots = \<T_{ij}\>_d, \\
   e^{\k\sub{0}} (2\pi R_y)^{2\s-d} \<T_{ab}\>_{2\s} &=& 2\s L e^{\k\sub{0}} g_{(2\s)ab} + \ldots \nono \\
   &=& 2\s L e^{\k\sub{0}} \left(e^{2\k/(2\s-d)}\right)_{(2\s)} \d_{ab} + \ldots \nono \\
   &=& \frac{4\s L}{2\s -d} e^{(1+2/(2\s-d))\k\sub{0}} \k\sub{2\s} \d_{ab} + \ldots \nono \\
   &=& -\<\cO_\phi\>_d e^{2\k\sub{0}/(2\s-d)} \d_{ab}, \nono
\eea
where the dots again contain curvatures of the boundary metric $g_{(0)ij}$ and derivatives of $\k_{(0)}$ and we used in the last line the results of \cite{Kanitscheider:2008kd} for the vev of the scalar operator,
\be
   \< \cO_\phi \>_d = - \frac{4\s L}{2\s-d} e^{\k\sub{0}} \k\sub{2\s} + \ldots.
\ee
The conformal Ward identity $\<T^\m_\m\>_{2\s} = \cA_{2\s}$ then reduces to
\bea
   && e^{-\k\sub{0}} (2\pi R_y)^{d-2\s} \left(\<T^i_i\>_{2\s} + g_{(0)}^{ab} \<T_{ab}\>_{2\s}\right) = \<T^i_i\>_d - (2\s-d) \<\cO_\phi\>_d \nono \\
   && = e^{-\k\sub{0}} (2\pi R_y)^{d-2\s}\cA_{2\s} \equiv \cA_d,
\eea
which is indeed equal to \eqref{Wardid}. The most efficient way to incorporate all local terms in the analysis
(which are denoted by dots here) is to use the Hamiltonian formulation of holographic renormalization \cite{Papadimitriou:2004ap,Papadimitriou:2004rz} and dimensionally reduce the results. This has been
discussed in detail for the case of D4 brane (which is related to M5 by the M-theory lift) in
 \cite{Kanitscheider:2008kd}.

Thus we find that local counterterms, 1-point functions and the
generalized conformal Ward identity for half-integer $\s > d/2$ can be obtained by dimensional reduction. From the lower-dimensional point of view however, $\s$ is just a parameter of the theory on which the equations of motion depend smoothly. Therefore local counterterms, 1-point functions and generalized conformal Ward identities should also exist for positive, but non-integer $\s >d/2$.

The reduction argument yields the following prescription to obtain the counterterms to \eqref{lowdaction} with $\s>d/2$ from $AdS$-counterterms. Choose any half-integer $\tilde{\s}> \s $ and determine the $[\s]+1$ most singular $AdS_{(2\tilde{\s}+1)}$-counterterms as a function of $\tilde{\s}$, where $[\s]$ denotes the largest integer less than or equal to $\s$ (when $\s$ is an integer one of these counterterms is logarithmic) . Reducing these $AdS_{(2\tilde{\s}+1)}$-counterterms on a $(2\tilde{\s}-d)$-dimensional torus and replacing $\tilde{\s}$ by $\s$ yields the counterterms appropriate for \eqref{lowdaction}.

As an example we rederive the counterterm action found in \cite{Kanitscheider:2008kd} for $1 < \s < 2$, which encompasses the cases of D0/1/2 branes and of the fundamental string, for which $\s = \{7/5, 3/2, 5/3,3/2\}$ and $d=\{1,2,3,2\}$ respectively. Since $\s < 2$ we only need two counterterms. The two most singular counterterms in $AdS_{2\tilde{\s}+1}$ defined on a regulating hypersurface are given by (see appendix B of \cite{deHaro:2000wj})\footnote{Note that convention for the curvature tensor used in \cite{deHaro:2000wj} has the opposite sign.}
\be
\label{Sct2sigma}
   S^{ct} = L_{AdS} \int_{\r=\e} d^{2\tilde{\s}}x \sqrt{-\g_{2\tilde{\s}}} \left[2(2\tilde{\s}-1) + \frac{1}{2\tilde{\s}-2} \hat{R}[\g_{2\tilde{\s}}]\right],
\ee
where $\g_{2\tilde{\s} ij}$ is the induced metric on the $(2\tilde{\s})$-dimensional hypersurface and $\hat{R}[\g_{2\tilde{\s}}]$ the corresponding curvature. The curvature on the hypersurface reduces to $d$ dimensions as
\be
  \hat{R}_{2\tilde{\s}} = \hat{R}_{d} [\g] - 2 \hna^2 \phi - \frac{2\tilde{\s}-d+1}{2\tilde{\s}-d} (\pa_i \phi)^2.
\ee
The counterterm action to \eqref{lowdaction} for $1< \s < 2$ is then given by reducing \eqref{Sct2sigma} to $d$ dimensions and replacing $\tilde{\s}$ with $\s$,
\be
   S^{ct} = L \int_{\r=\e} d^dx \sqrt {-\g_d}\,\, e^\phi \left[2 (2\s-1) + \frac{1}{2\s-2} (\hat{R}_d + \frac{2\s -d -1}{2\s-d} (\pa_i \phi)^2)\right],
\ee
which agrees with formula (5.74) of \cite{Kanitscheider:2008kd}. The remaining case, i.e. the case of D4 branes
has $\s=3$ and the counterterm action also follows in the same manner (i.e. from the gravitational
counterterms for $AdS_7$), as discussed in detail
in \cite{Kanitscheider:2008kd}. Finally, let us comment on the restriction $\s > d/2$.
At $\s =d/2$ the action (\ref{Ricactred}) has a pole and the kinetic
term of the scalar becomes negative in the interval $(d-1)/2 < \s < d/2$ so one should use the reduction
argument when $\s<d/2$ with caution. Note also that for D6 branes, which do not have a sensible
decoupling limit, $\s=-1$.

For later convenience let us finally mention that one can always formally recover the $(2\s+1)$-dimensional equation of motion for the metric in the conformal case from the non-conformal case by setting the scalar to zero. The conformal version of \eqref{tra-eq} - \eqref{Ricasympt} reads
\bea
   &&  -\frac{1}{4} \Tr(g^{-1}g')^2 + \hp \Tr g^{-1} g'' = 0, \label{conf-tra-eq} \\
&&  -\hp\na^\m {g'}_{\m\n} + \hp \pa_\n
(\Tr g^{-1} g') = 0, \label{conf-div-eq} \\
&& \left[-Ric(g) -2(\s-1)g' - \Tr(g^{-1}g')g +
  \r(2g''
- 2g'g^{-1}g'+ \Tr(g^{-1}g')g')\right]_{\m\n}  = 0, \qquad  \label{conf-Ricasympt}
\eea
where from now on we use transverse indices $\m,\n,\ldots$ for the conformal case and transverse indices $i,j,\ldots$ for the non-conformal case.

\section{Universal Hydrodynamics} \label{three}
\label{univhydro}

The hydrodynamic energy-momentum tensor for a conformal fluid at first-derivative order in $(2\s)$ dimensions on a curved manifold with metric $g\sub{0}_{\m\n}$ is
\bea
\label{hydroconfTmn}
   T_{\m\n} &=& L_{AdS}\left(\frac{2\pi T}{\s}\right)^{2\s}(g\sub{0}_{\m\n} + 2\s u_\m u_\n) - 2\eta_{2\s}(T) \s_{\m\n}, \\
   \s_{\m\n} &=& P_\m^\k P_\n^\l \na_{(\k} u_{\l)} - \frac{1}{2\s-1} P_{\m\n} (\na \cdot u), \qquad P_{\m\n} = g\sub{0}_{\m\n} + u_\m u_\n, \nono
\eea
where $T$, $u_\m$ and $\eta_{2\s}(T)$ denote the temperature, velocity and shear viscosity respectively of the fluid and $\na_i$ is the covariant derivative corresponding to the metric $g\sub{0}_{ij}$. For given $\eta_{2\s}(T)$ the evolution of the fluid is determined by the conservation of the energy-momentum tensor,
\be
\label{convcons}
   \na^\m T_{\m\n} = 0.
\ee
Furthermore, the conformal Ward identity $T^\m_\m =0$ constrains energy density $\e$ and pressure $p$ to be related by the equation of state
\be
\label{eostate}
   p = L_{AdS}\left(\frac{2\pi T}{\s}\right)^{2\s} = \frac{1}{2\s-1} \e.
\ee
Since we saw above that the bulk equations of motion for a non-conformal geometry with given $\s$ can be obtained by dimensional reduction of $(2\s+1)$-dimensional gravity on a $(2\s-d)$-dimensional torus, we can perform the same procedure on the boundary to obtain the hydrodynamic energy-momentum tensor dual to a non-conformal black brane solution with given $\s$ in $d$ dimensions. Demanding that $T_{\m\n}$ in \eqref{hydroconfTmn} only depends on non-compact directions and that the fluid velocity $u^\m = (u^i, 0)$ only has non-zero non-compact components yields
\bea
\label{hydrosigmaTij}
   T_{ij} &=& L e^{\k\sub{0}}\left(\frac{2\pi T}{\s}\right)^{2\s}(g\sub{0}_{ij} + 2\s u_iu_j)  - 2\eta_d \s_{ij} - \zeta_d P_{ij} (\na \cdot u), \\
   \<\cO_\phi\> &=& -L e^{\k\sub{0}}\left(\frac{2\pi T}{\s}\right)^{2\s} - \frac{2}{2\s-1}\eta_d (\na \cdot u), \nono
\eea
where
\bea
\label{hydrosigmaTij2}
   \s_{ij} &=& P_i^kP_j^l \na_{(k} u_{l)} - \frac{1}{d-1} P_{ij} (\na \cdot u), \qquad P_{ij} = g\sub{0}_{ij} + u_iu_j, \\
   \eta_d &=& (2\pi R_y)^{2\s-d} e^{\k\sub{0}} \eta_{2\s}, \nono \\
   \zeta_d &=& \frac{2(2\s -d)}{(d-1)(2\s-1)} \eta_d, \nono
\eea
with $\eta_d$ and $\zeta_d$ shear and bulk viscosity respectively of the $d$-dimensional fluid. The conformal conservation equation \eqref{convcons} reduces to \cite{Kanitscheider:2008kd}
\be
   \na^i T_{ij} - \pa_j \k\sub{0} \< \cO_\phi \> = 0,
\ee
where $\< \cO_\phi \> $ is again the expectation value of the operator dual to $\phi$. Since we would like the evolution of the $d$-dimensional fluid to be described purely by a divergence equation, we demand that $\k\sub{0}$ is constant or without loss of generality zero. Moreover note that the $d$-dimensional non-conformal fluid obeys the same equation of state \eqref{eostate} as the $(2\s)$-dimensional conformal fluid.

In \eqref{hydrosigmaTij2} we observe that $\eta_d(T)$ as function of the temperature in the non-conformal theory is proportional to $\eta_{2\s}(T)$ in the higher dimensional conformal theory. As we will check below, smoothness of the bulk solution forces the viscosity to saturate the KSS bound \cite{KSSbound}
\be
\label{etasbound}
  \frac{\eta_d}{s_d} = \frac{\eta_{2\s}}{s_{2\s}} \geq \frac{1}{4\pi},
\ee
where the entropy density $s_d$ corresponding to \eqref{hydrosigmaTij} is given by
\be
\label{entrdeptemp}
   s_d = 2\s L \left(\frac{2\pi}{\s}\right)^{2\s} T^{2\s-1},
\ee
and the dimensionful $L$ is related to the dimensionless $L_{AdS}$ via \eqref{llads}. More generally, we see that any fluid which is related to a conformal fluid satisfying the KSS bound by dimensional reduction will satisfy the KSS bound as well.

Furthermore we note that in \eqref{hydrosigmaTij2}, the bulk viscosity $\zeta_d$ is determined by the shear viscosity $\eta_d$. In \cite{Buchel:2007mf} it was conjectured that the ratio of bulk to shear viscosity of a strongly coupled gauge theory plasma satisfies the bound\footnote{Note that unlike the $\eta/s$ bound, this bound has known counterexamples in weakly coupled systems, e.g. monatomic gases \cite{Buchel:2007mf}.}
\be
\label{bvbound}
   \frac{\zeta_d}{\eta_d} \geq 2(\frac{1}{d-1} - c_s^2),
\ee
where $c_s$ denotes the speed of sound. In our case $c_s$ can be calculated from the equation of state \eqref{eostate} to be
\be
\label{ssound}
   c_s = \sqrt{\frac{\pa p}{\pa \e}} = \frac{1}{\sqrt{2\s-1}}.
\ee
Hence we see that the last line in \eqref{hydrosigmaTij2} implies that the bound \eqref{bvbound} is saturated for arbitrary $\s$. This confirms the calculation of \cite{Mas:2007ng, Benincasa:2006ei, Buchel:2007mf}, in which this result was obtained for black D$p$-branes for $p=2,\ldots,6$ and their toroidal compactifications. Note however that \eqref{bvbound} will be saturated for any fluid which arises from dimensional reduction
and continuation in dimension of a conformal fluid, irrespective of the value of $\eta/s$. In particular we did not have to assume that the dual bulk solution is smooth. Thus in this case the ratio $\zeta/\eta$ is fixed
kinematically. This indicates that this case is qualitatively
different than the case of $\eta/s$.

\comment{By combining these ingredients we find a bound for the ratio of bulk viscosity to entropy,
\be
   \frac{\zeta_d}{s_d} \geq \frac{1}{2\pi}(\frac{1}{d-1} - c_s^2).
\ee}
Combining \eqref{hydrosigmaTij2} and \eqref{etasbound} we obtain the bound
\be
   \frac{\zeta_d}{s_d} \geq \frac{2\s-d}{2\pi (d-1)(2\s-1)}.
\ee
which should hold for all non-conformal fluids that can be related to a $(2 \s)$-dimensional
conformal fluid (with $(2 \s)$ non necessarily integral), as discussed above.

We can also obtain to second order the coefficients of the non-conformal energy-momentum tensor from the coefficients of the conformal energy-momentum tensor.
In was argued in \cite{Baier:2007ix} that the second order contribution to the conformal energy-momentum tensor is given by a linear combination of all possible Weyl invariants containing two derivatives,
\bea
\label{confTij2nd}
   T_{2\m\n} &=& 2\eta_{2\s} \tau_M \left[ (u\cdot \na) \s_{\m\n} + \frac{1}{2\s-1} \s_{\m\n} (\na \cdot u)\right] \\
  && + \tilde{\k} \left[ R_{\m\n} - (2\s-2) u^\k u^\l R_{\k \<\m\n\> \l} \right] \nono \\
  && +4\l_1 \s_{\k\<\m} {\s_{\n\>}}^\k + 2 \l_2 \s_{\k\<\m} {\W_{\n\>}}^\k + \l_3 \W_{\k\<\m} {\W_{\n\>}}^\k,  \nono
\eea
where $R_{\m\n\k\l}$ and $R_{\m\n}$ are Riemann and Ricci tensor of the metric $g\sub{0}_{\m\n}$, angle brackets denote the transverse traceless part of a second rank tensor $A_{\m\n}$,
\be
\label{ttpart2sigma}
   A_{\<\m\n\>} = \hp P_\m^\k P_\n^\l ( A_{\k\l} + A_{\l\k}) - \frac{1}{2\s-1} P_{\m\n} P^{\k\l} A_{\k\l},
\ee
and the vorticity $\W_{\m\n}$ is given by
\be
   \W_{\m\n} =  \hp P_\m^\k P_\n^\l (\na_\k u_\l - \na_\l u_\k).
\ee
Note also that with notation \eqref{ttpart2sigma} the shear tensor $\s_{\m\n}$ can be written as
\be
   \s_{\m\n} = \na_{\<\m} u_{\n\>}.
\ee
Again, we can obtain the non-conformal second order energy-momentum tensor with given $\s$ by reducing \eqref{confTij2nd} on a $(2\s-d)$-dimensional torus. The result can be obtained by replacing all tensors with angle brackets by
\bea
   A_{\<\m\n\>} &\rightarrow& A_{\<ij\>} + \frac{2\s-d}{(d-1)(2\s-1)} P_{ij} P^{kl} A_{kl}, \\
   \s_{\m\n} &\rightarrow& \s_{ij} + \frac{2\s-d}{(d-1)(2\s-1)} P_{ij} (\na \cdot u), \nono
\eea
where $A_{\<ij\>}$ in the first line on the right hand side is defined as transverse traceless part in the lower dimensional theory,
\be
\label{ttpart2d}
   A_{\<ij\>} = \hp P_i^kP_j^l ( A_{kl} + A_{lk}) - \frac{1}{d-1} P_{ij} P^{kl} A_{kl}.
\ee
The Riemann tensor, Ricci tensor, Ricci scalar and vorticity reduce trivially,
\be
   R^{2\s}_{ijkl} = R^d_{ijkl}, \qquad R^{2\s}_{ij} = R^d_{ij}, \qquad R^{2\s} = R^d, \qquad \W^{2\s}_{ij} = \W^{d}_{ij},
\ee
since we demanded that $\k\sub{0} = 0$. The components of the Riemann and Ricci tensor in the internal directions of the torus do not contribute to the $d$-dimensional energy-momentum tensor. Finally the $d$-dimensional energy momentum tensor gets multiplied by the overall factor $(2\pi R_y)^{2\s-d}$ which stems from the torus volume factor multiplying the $(d+1)$-dimensional bulk action.

\section{Generalized black branes} \label{five}

In \cite{Bhattacharyya:2008jc} it was shown that the long wavelength fluctuation equations around the boosted black D3 brane geometry in Eddington-Finkelstein coordinates can be mapped to the non-linear equations of hydrodynamics of the dual strongly coupled conformal field theory.
In \cite{Gupta:2008th} it was pointed out that it can be advantageous to perform the same analysis in Fefferman-Graham coordinates, since it allows for a Lorentz covariant expansion, the constraint equations become trivial and reading off the stress tensor from the bulk metric is completely straightforward. Furthermore,
one can construct bulk solutions dual to an arbitrary hydrodynamic boundary stress tensor. On the other hand, irrespectively of the precise values of the coefficients of the energy-momentum tensor, the Fefferman-Graham coordinates will have a singularity at the (unperturbed) horizon. To find out whether this singularity is a coordinate singularity or a real one requires to transform to Eddington-Finkelstein coordinates. Only requiring smoothness in Eddington-Finkelstein coordinates away from the singularity of the static black brane fixes the coefficients in the boundary stress tensor to the values found in \cite{Bhattacharyya:2008jc}.

Here we will generalize the analysis of \cite{Gupta:2008th} to non-conformal geometries with $AdS$-solution in the dual frame and arbitrary positive $\s$. As a first step, we generalize it to pure gravity in arbitrary dimension $(2\s+1)$ for half-integer $\s$ and then invoke the reduction argument of section \ref{lowdfe} to obtain the case of a non-conformal geometry with arbitrary positive $\s$ in dimension $d$. Throughout the rest of the paper we assume the boundary metric $g_{(0)ij}=\eta_{ij}$ to be flat and $\k\sub{0}$ to be constant or without loss of generality zero.

For half-integer $\s$, the $(2\s+1)$-dimensional pure gravity action in \eqref{Ricactred} has the black brane solution
\bea
\label{bbraneconfdt}
   ds^2 &=& \frac{d\r^2}{4\r^2f_b(\r)} + \frac{ -f_b(\r)dt^2 + dz_r dz^r}{\r}, \\
   f_b(\r) &=& 1 - \frac{\r^\s}{b^{2\s}}, \nono
\eea
where $r$ runs over spatial transverse coordinates and $b$ is related to the black brane temperature by
\be
\label{btemp}
   b = \frac{\s}{2\pi T}.
\ee
After boosting the geometry \eqref{bbraneconfdt} with the boost parameter $u_\m$ we obtain the metric
\bea
\label{boostbbrane}
   ds^2 &=& \frac{d\r^2}{4\r^2f_b(\r)} + \frac{ [\eta_{\m\n} + (1-f_b(\r))u_\m u_\n] dz^\m dz^\n}{\r},
\eea
which solves the equation of motions as long as $b$ and $u_\m$ are constants, with $b$ and $u_\m$ mapped to the to the dual (inverse) temperature and velocity of the fluid. However, once we allow the temperature in the definition of $b$ in \eqref{btemp} and $u_\m$ to become $z$-dependent,
\bea
\label{boostbbranedep}
   ds^2 &=& \frac{d\r^2}{4\r^2f_{b(z)}(\r)} + \frac{ [\eta_{\m\n} + (1-f_{b(z)}(\r))u_\m (z)u_\n(z)] dz^\m dz^\n}{\r},
\eea
 we have to correct the metric \eqref{boostbbranedep} at each order in the derivative expansion to still fulfill the equations of motions. The corrections to the metric then determine the dissipative part of the hydrodynamic energy-momentum tensor.

 The non-conformal generalization of \eqref{boostbbranedep} can again be obtained by compactification. We split the transverse coordinates $z^\m = (z^i, y^a)$ in non-compact and torus directions and demand that the metric only depends on non-compact directions and that the fluid velocity $u^\m = (u^i,0)$ has only non-zero non-compact components. This enables us to reduce using the reduction ansatz \eqref{redans} to obtain for metric and scalar
\bea
\label{bbranedt}
  ds^2 &=& \frac{d\r^2}{4\r^2f_b(\r)} + \frac{[\eta_{ij} + (1-f_{b(z)}(\r))u_i (z)u_j(z)] dz^i dz^j}{\r}, \\
  e^\phi &=& \r^{\a\g}. \nono
\eea
It can be checked explicitly that this is a solution of the equations of motion following from the $(d+1)$-dimensional action \eqref{lowdaction} for arbitrary $\s$.

\section{Generalized black branes in Fefferman-Graham coordinates} \label{six}
\label{GenbbFG}

Before computing the derivative corrections to the boosted brane solution \eqref{bbranedt} by perturbing around equations \eqref{tra-eq} - \eqref{scalasympt}, we change to Fefferman-Graham coordinates, in which the solution takes the form \eqref{metrscalans}. Again, to keep the discussion as concise as possible, we first discuss the conformal case in arbitrary dimension and compactify to obtain the non-conformal case. In both cases, we obtain Fefferman-Graham coordinates by a redefinition of the radial coordinate:
\be
   \rt(\r) = \left(\frac{2}{1+\sqrt{f_b(\r)}}\right)^{2/\s} \r,
\ee
whose inverse transformation is
\be
   \r(\rt) = \left(1+ \frac{\rt^\s}{4 b^{2\s}}\right)^{-2/\s} \rt.
\ee
The metric \eqref{boostbbranedep} corresponding to the conformal fluid becomes
\bea
\label{bbranefg}
   ds^2 &=& \frac{d\rt^2}{4\rt^2} + \frac{g(z,\rt)_{\m\n}dz^\m dz^\n}{\rt}, \\
   g(z,\rt)_{\m\n} &=& A(\rt) \eta_{\m\n} + B(\rt) u_\m u_\n, \nono
\eea
where
\bea
   A(\rt) &=& \frac{\rt}{\r(\rt)} = \left(1+ \frac{\rt^\s}{4 b^{2\s}}\right)^{2/\s}, \\
   B(\rt) &=& \frac{\rt[1-f_b(\r(\rt))]}{\r(\rt)} = \frac{\rt^\s}{b^{2\s}}\left(1+\frac{\rt^\s}{4b^{2\s}}\right)^{2/\s-2}. \nono
\eea
According to \eqref{conf_dict} we obtain the perfect fluid part of the energy-momentum tensor \eqref{hydroconfTmn} by reading off the $\rt^\s$ coefficient of $g(z,\rt)$,
\be
\label{pfTij}
   T_{0\m\n} = 2\s L_{AdS} g\sub{2\s}_{\m\n} = L_{AdS} b^{-2\s} (\eta_{\m\n} + 2\s u_\m u_\n),
\ee
using the definition of $b$ in \eqref{btemp}. The horizon in Fefferman-Graham coordinates is at $\rt = \rt_h \equiv 2^{2/\s}b^2$, where $g(\rt_h,z)_{\m\n}$ becomes non-invertible since $A(\rt_h)=B(\rt_h)$.

If $u_\m(z)$ and $b(z)$ in \eqref{bbranefg} become dependent on the boundary coordinates $z^\m$ we have to introduce corrections to the metric at each order in the derivative expansion to still satisfy the equations of motion. At first order we perturb the metric as
\be
   g(z,\rt) = g_0(z,\rt) + g_1(z,\rt),
\ee
where $g_0(z,\rt)$ is given by \eqref{bbranefg},
\be
   g_0(z,\rt)_{ij} = A(b(z),\rt) \eta_{ij} + B(b(z),\rt) u_i(z) u_j(z).
\ee
The equations of motion \eqref{conf-tra-eq} and \eqref{conf-Ricasympt} become
\bea
  && -\hp \Tr g_0^{-1} g_0' g_0^{-1} g_1' + \hp \Tr g_0^{-1} g_1 g_0^{-1} g_0' g_0^{-1} g_0' \label{conf-trapert}
  + \hp(\Tr g_0^{-1} g_1'' - \Tr g_0^{-1} g_1 g_0^{-1} g_0'') = 0,
  \\
  &&  2\rt (g_1'' - g_1'g_0^{-1}g_0' - g_0'g_0^{-1}g_1' + g_0'g_0^{-1}g_1g_0^{-1}g_0') -2(\s-1) g_1' \label{conf-Ricpert} \\
  && \qquad +\Tr g_0^{-1} g_0'(\rt g_1' - g_1) + (\Tr g_0^{-1} g_1' - \Tr g_0^{-1} g_1 g_0^{-1} g_0')(\rt g_0' - g_0) = 0, \nono
\eea
where now the prime denotes differentiation with respect to the Fefferman-Graham radial variable $\rt$. At first order in the derivative expansion, $Ric(g)$ in \eqref{conf-Ricasympt} does not contribute since it contains at least two derivatives of $z$.

At every order in the derivative expansion, the constraint equation \eqref{conf-div-eq} at small $\rt$ is equivalent to the conservation of the dual energy-momentum tensor. However, if this equation is fulfilled on a radial hypersurface close to the boundary, the evolution equations \eqref{conf-tra-eq} and \eqref{conf-Ricasympt} ensure that it remains fulfilled in the interior. Only the equations \eqref{conf-trapert} and \eqref{conf-Ricpert} constrain the form of the metric perturbation further and with it also the form of the dual hydrodynamic stress tensor. In section \ref{trafoEF} though, where we transform the perturbed metric to Eddington-Finkelstein coordinates, it will be convenient to use the conservation equation of the (perfect fluid) energy-momentum tensor to relate derivatives of the temperature field to derivatives of the velocity field.

 The perturbations $g_{1\m\n}$ will contain first derivatives of $u_\m$ and its order $\s$ term will correct the energy momentum tensor by  $T_{\m\n} = T_{0\m\n} + T_{1\m\n}$, where $T_{0\m\n}$ is the perfect fluid energy-momentum tensor \eqref{pfTij} and $T_{1\m\n}$ the dissipative part at first derivative order. Since $T^\m_\m =0$ in the conformal case and since we can always go to Landau gauge $u^\m T_{1\m\n} = 0$ by a redefinition of the temperature and velocity field, $T_{1\m\n}$ will be given by
\be
   T_{1\m\n} = -2\eta_{2\s} \s_{\m\n},
\ee
where the parameter $\eta_{2\s}$ is the shear viscosity. Only for a specific value of the shear viscosity, the bulk solution will be smooth at the horizon of the black brane. However, in Fefferman-Graham coordinates the metric becomes non-invertible at the horizon. Fixing the value of $\eta_{2\s}$ will require changing to Eddington-Finkelstein coordinates, which we do in section \ref{trafoEF} below. In the meantime we parametrize $\eta_{2\s}$ as
\be
   \eta_{2\s} = L_{AdS}\g b^{1-2\s},
\ee
where $\eta_{2\s}$ fulfilling $\eta_{2\s}/s_{2\s} = 1/4\pi$ corresponds to $\g=1$.

The form of the metric perturbation $g_{1\m\n}$ can now be determined using the
the following argument \cite{Gupta:2008th}. As is shown in appendix \ref{appA}, the derivatives in the $\rt$ expansion of the metric  always enter in pairs, see \eqref{gsub2tau}, which implies that the $\rt$-expansion of the metric perturbation $g_{1\m\n}$ at first derivative order will only contain non-derivative terms of the form $(T_0^pT_1T_0^q)$. Due to the Landau gauge condition $u^\m T_{1\m\n} =0$ and the tracelessness condition $T^\m_{1\m} =0$ only the $\eta_{\m\n}$ part inside $(T_0^p)_{\m\n}$ contributes to the coefficients of $g_{1\m\n}$ at each order in $\rt$. Thus, each coefficient in the expansion of $g_{1\m\n}$ will be proportional to $T_{1\m\n}$. Hence also $g_{1\m\n}$ as a whole will be proportional to $T_{1\m\n}$, which in the conformal case only contains a shear part,
\be
\label{FG1orderconf}
   g_{1\m\n} = \l(\rt) \s_{\m\n}.
\ee
\comment{
In the nonconformal case, the metric and scalar perturbation $g_{1ij}$ and $\k_1$ will be a linear combination of first order differentials of $u_i$.
The most general ansatz is
\bea
\label{FG1order}
   g_{1ij} &=& \l(\rt) \s_{ij} + \mu(\rt) P_{ij} (\pa \cdot u) + \w_1(\rt) u_iu_j (\pa \cdot u) + \w_2(\rt) (u\cdot \pa) (u_iu_j), \\
   \k_1 &=& \nu(\rt) (\pa \cdot u). \nono
\eea}
Extracting the transverse, traceless mode proportional to $\s_{\m\n}$ out of \eqref{conf-Ricpert} we obtain a second order ordinary differential equation in $\l(\rt)$
\be
   2\rt(\l''-2\frac{A'}{A}\l'+\frac{A'^2}{A^2} \l) - 2(\s-1)\l'+\Tr g_0^{-1}g_0'(\rt\l' - \l) = 0,
\ee
whose asymptotically vanishing solution is given by
\be
   \l(\rt) = C_\l \left(1+\frac{\rt^\s}{4 b^{2\s}}\right)^{2/\s} \log \frac{1-\frac{\rt^\s}{4 b^{2\s}}}{1+\frac{\rt^\s}{4 b^{2\s}}} = C_\l A(\rt) \log \frac{2-A(\rt)^{\s/2}}{A(\rt)^{\s/2}}.
\ee
To fix the integration constant $C_\l$ we demand that the order $\s$ term of $\l(\rt)$ in the Taylor expansion in $\rt$ reproduces $T_{1\m\n}$,
\be
   C_\l = \frac{2\eta_{2\s}}{ \s L_{AdS}} b^{2\s} = \frac{2\g b}{\s}.
\ee
\comment{
Moreover, we argue that $\w_2(\rt)=0$ by the following argument. Since $T_{1ij}$ is in Landau gauge, it does not contain a term proportional to $(u\cdot \pa) (u_iu_j)$. Assuming $\w_2(\rt) \neq 0$, let $\w_2(\rt)$ have the small $\rt$ expansion
\be
\label{w2expansion}
   \w_2(\rt) = \w_{2(2\tau)} \rt^{\tau} + subleading, \qquad \w_{2(2\tau)} \neq 0, \qquad  \tau > \s.
\ee
Extracting the mode proportional to $(u\cdot \pa) (u_iu_j)$ out of \eqref{Ricpert} we obtain after expanding near $\rt=0$ the differential equation
\be
\label{w2diffequ}
   2\rt \w_2'' + \w_2' (-2(\s-1) + \cO(\rt^\s)) + \w_2 \cO(\rt^{2\s-2}) =0.
\ee
Plugging in the ansatz \eqref{w2expansion} in \eqref{w2diffequ} we obtain for the order $\rt^{\tau-1}$ coefficient
\be
   2 \tau(\tau-\s) \w_{2(2\tau)} = 0.
\ee
Due to the assumption $\tau > \s$, it follows $\w_{2(2\tau)}=0$ in contradiction with above. Therefore $\w_2(\rt)$ has to vanish identically.

Equations \eqref{trapert}, \eqref{Ricpert} and \eqref{scalpert} then yield a coupled second order differential equation for $\mu(\rt), \nu(\rt)$ and $\w_1(\rt)$. Similarly to the discussion in section \ref{univhydro}, we can find a solution by reduction from the conformal case in $(2\s+1)$ bulk dimensions, where only $\l(\rt)$ is non-vanishing. For metric and scalar at zeroth and first order we obtain in this way
\bea
   g_{ij} &=& A \eta_{ij} + B u_i u_j + \l \s_{ij} + \frac{2\s-d}{(d-1)(2\s-1)} \l P_{ij} (\pa \cdot u), \nono \\
   \exp(\frac{2\k}{2\s-d}) &=& A- \frac{\l}{2\s-1} (\pa \cdot u),
\eea
hence,
\be
\label{exprmunu}
   \mu = \frac{2\s-d}{(d-1)(2\s-1)} \l, \qquad \nu = -\frac{2\s-d}{2(2\s-1)A} \l, \qquad \w_1=0.
\ee
It is straightforward to check that \eqref{exprmunu} solves equations \eqref{Ricpert} and \eqref{scalpert} for $\w_1=\w_2 =0$,
\bea
   &&-\hp \frac{(d-1)A'}{A^2}  \mu' + \hp \frac{(d-1)A'^2}{A^3}  \mu + \hp \frac{(d-1)}{A}\mu'' - \hp\frac{(d-1)A''}{A^2} \mu \\
   && \qquad  + \nu'' - \frac{1}{\a\g} \k_0' \nu' =0, \nono \\
   && 4 \rt (\nu'' + 2 \k_0' \nu') + (2d+4-8\s) \nu' + (d-2\s+2\rt\k_0')\left(\frac{(d-1)}{A}\mu' - \frac{A'}{A^2} (d-1) \mu \right) \nono \\
   && \qquad   + 2\rt \nu' \Tr(g_0^{-1} g_0') =0.
\eea
}
The metric in Fefferman-Graham coordinates in the conformal case up to first order is then
\bea
\label{fullFG1orderconf}
   ds^2 &=& \frac{d\rt^2}{4\rt^2}+ \frac{g_{\m\n}(z,\rt) dz^\m dz^\n}{\rt}, \\
   g_{\m\n}(z,\rt) &=& A(\rt) \eta_{\m\n} + B(\rt) u_\m u_\n + \l(\rt) \s_{\m\n}. \nono
\eea
The whole discussion can be straightforwardly generalized to the nonconformal case. Starting from \eqref{bbranedt} we change to Fefferman-Graham coordinates, perturb metric and scalar as
\bea
   g(z,\rt) &=& g_0(z,\rt) + g_1(z,\rt),  \\
   \k(z,\rt) &=& \k_0(z,\rt) + \k_1(z,\rt), \nono
\eea
to obtain perturbation equations around \eqref{tra-eq}, \eqref{Ricasympt} and \eqref{scalasympt}:
\bea
  && -\hp \Tr g_0^{-1} g_0' g_0^{-1} g_1' + \hp \Tr g_0^{-1} g_1 g_0^{-1} g_0' g_0^{-1} g_0' \label{trapert} \\
  && \qquad + \hp(\Tr g_0^{-1} g_1'' - \Tr g_0^{-1} g_1 g_0^{-1} g_0'') + \k_1'' - \frac{1}{\a\g} \k_0' \k_1' = 0, \nono \\
  &&  2\rt (g_1'' - g_1'g_0^{-1}g_0' - g_0'g_0^{-1}g_1' + g_0'g_0^{-1}g_1g_0^{-1}g_0') -2(\s-1) g_1' \label{Ricpert} \\
  && \qquad +(\Tr g_0^{-1} g_0' + 2\k_0') (\rt g_1' - g_1) \nono \\
  && \qquad + (\Tr g_0^{-1} g_1' - \Tr g_0^{-1} g_1 g_0^{-1} g_0' + 2\k_1')(\rt g_0' - g_0) = 0 \nono \\
  && 4\rt (\k_1'' + 2\k_1' \k_0') + 2(d-4\s+2) \k_1'  \label{scalpert} \\
  && \qquad + (\Tr g_0^{-1} g_1' - \Tr g_0^{-1} g_1 g_0^{-1} g_0')(2\a\g + 2\rt \k_0')
  + 2\rt \k_1'\Tr g_0^{-1} g_0'  = 0. \nono
\eea
However, as we know the first order solution \eqref{fullFG1orderconf} in the conformal case, we can again obtain the first order solution for metric and scalar in the non-conformal case by dimensional reduction using the ansatz \eqref{redans}:
\bea
\label{fullFG1ordernonconf}
   g_{ij}(z,\rt) &=& A(\rt) \eta_{ij} + B(\rt) u_i u_j + \l(\rt) \s_{ij} + \frac{2\s-d}{(d-1)(2\s-1)} \l(\rt) P_{ij} (\pa \cdot u), \nono \\
   \exp(\frac{2\k(z,\rt)}{2\s-d}) &=& A(\rt)- \frac{\l(\rt)}{2\s-1} (\pa \cdot u).
\eea
From \eqref{fullFG1ordernonconf} we can read off
\bea
   g_0(z,\rt)_{ij} &=& A(\rt) \eta_{ij} + B(\rt) u_i(z) u_j(z), \\
   \k_0(z,\rt) &=& \frac{2\s - d}{2} \log A(\rt), \nono \\
   g_1(z,\rt)_{ij} &=& \l(\rt) \s_{ij} + \frac{2\s-d}{(d-1)(2\s-1)} \l(\rt) P_{ij} (\pa \cdot u), \nono \\
   \k_1(z,\rt) &=& -\frac{2\s-d}{2(2\s-1)A(\rt)} \l(\rt) (\pa \cdot u), \nono
\eea
which can be straightforwardly shown to be a solution of \eqref{trapert} - \eqref{scalpert}. Finally, by extracting the order $\s$ term, it can be checked that the metric in \eqref{fullFG1ordernonconf} gives rise to the non-conformal energy momentum tensor and scalar vev given in \eqref{hydrosigmaTij},
\bea
   T_{ij} &=& \frac{L}{b^{2\s}}(g\sub{0}_{ij} + 2\s u_iu_j)  - 2\eta_d \s_{ij} - \zeta_d P_{ij} (\pa \cdot u), \\
   \<\cO_\phi\> &=& -\frac{L}{b^{2\s}} - \frac{2}{2\s-1}\eta_d (\pa \cdot u). \nono
\eea

\section{Transformation to Eddington-Finkelstein coordinates} \label{seven}
\label{trafoEF}

We have thus found that the first order perturbation results in a hydrodynamic stress energy tensor
and vev for the operator $\cO_\phi$ that are parametrized by the shear viscosity $\eta_d$,
which at this point is unconstrained. The bulk viscosity $\zeta_d$ is fixed in a way prescribed by
the dilation Ward identity. Now recall that the source and the vev are a conjugate pair
with the vev being the (renormalized) radial canonical momentum \cite{Papadimitriou:2004ap, Papadimitriou:2004rz} so specifying them yields in principle a unique bulk solution.
Not all these solutions however will be non-singular. Regularity in the interior in general
leads to additional restrictions.

We now discuss the constraints imposed by the smoothness of the gravity solution in the bulk.
This requires changing to Eddington-Finkelstein coordinates, which are well-defined beyond the horizon. In the conformal case the metric in the Eddington-Finkelstein coordinates will be of the form
\be
\label{efform}
   ds^2 = -2u_\m(x) dr dx^\m + G_{\m\n}(x,r) dx^\m dx^\n,
\ee
and the transformation equations for the metric between Fefferman-Graham of the form \eqref{metrscalans} and Eddington-Finkelstein coordinates are given by
\bea
\label{FGEFtrafo}
   && (\pa_r \rt)^2 + 4 \rt \, g_{\m\n}(z,\rt) \, \pa_r z^\m \, \pa_r z^\n = 0,  \\
   && \pa_r \rt\,  \pa_\m \rt + 4 \rt\,  \pa_r z^\k\, \pa_\m z^\l\, g_{\k\l}(z,\rt) = - 4\rt^2 u_\m, \nono \\
   && \pa_\m \rt\, \pa_\n \rt + 4 \rt\, \pa_\m z^\k\, \pa_\n z^\l\, g_{\k\l}(z,\rt) = 4 \rt^2 G_{\m\n}(x,\r), \nono
\eea
where $\rt(x,r)$ and $z^\m(x,r)$ encode the dependence of the Fefferman-Graham coordinates on the Eddington-Finkelstein coordinates. Given a solution in Fefferman-Graham coordinates we use \eqref{FGEFtrafo} to solve for $\rt(x,r)$, $z^\m(x,r)$ and $G_{\m\n}(x,\r)$.
At zeroth order in the derivative expansion, the transformation is given by
\bea
\label{0ordertrafo}
   \rt_0(r) &=& \rt(\rho=1/r^2) = \left(\frac{2}{1+\sqrt{f_b(r)}}\right)^{2/\s}\frac{1}{r^2}, \\
   z^\m_0(r) &=& x^\m + u^\m k_b(r), \nono
\eea
where
\be \label{hyper}
   f_b(r) \equiv f_b(\rho=1/r^2) = 1- (br)^{-2\s}, \qquad k_b(r) \equiv \frac{1}{r}\, {}_2F_1(1, \frac{1}{2\s}; 1+ \frac{1}{2\s}; (br)^{-2\s}),
\ee
with ${}_2F_1(a,b;c;w)$ a hypergeometric function. Note that the radial coordinate $r$ in the Eddington-Finkelstein coordinates is related to the radial coordinate $\r$ in the original black brane solution \eqref{bbranedt} simply by $\r = 1/r^2$. Furthermore $\rt(r)$ and $k_b(r)$ obey the first order differential equations
\bea
   \pa_r \rt &=& - \frac{2 \rt}{r \sqrt{f_b(r)}}, \nono \\
   \pa_r k_b &=& - \frac{1}{r^2 f_b(r)}.
\eea
$G_{0\m\n}(x,r)$ is given by
\be
\label{ef0order}
   G_{0\m\n} = r^2[\eta_{\m\n} + (f_b(r)-1)u_\m u_\n].
\ee
At first order in the derivative expansion we perturb the Fefferman-Graham coordinates in the transformation equations \eqref{FGEFtrafo} by
\bea
   \rt(x,r) &=& \rt_0(x,r) + \rt_1(x,r), \\
   z^\m(x,r) &=& z^\m_0(x,r) + z^\m_1(x,r), \nono
\eea
while $G_{\m\n}(x,r)$ is expanded as
\be
   G_{\m\n}(x,r) = G_{0\m\n}(x,r) + G_{1\m\n}(x,r).
\ee
At the same time, we use for the Fefferman-Graham metric the full first order expression $g_{\m\n}(z,\rt) = g_{0\m\n}(z,\rt) + g_{1\m\n}(z,\rt)$ in \eqref{fullFG1orderconf}. Note that the zeroth order expressions $\rt_0$, $z^\m_0$, $G_{0\m\n}$ and $g_{0\m\n}(z,\rt)$ depend also on $x$ through their dependence on $b(z)$ and $u_\m(z)$, which now have been made dependent on $z$. Furthermore, we have to account for the transverse coordinates change from $z^\m$ to $x^\m$ in \eqref{0ordertrafo} and Taylor expand $b(z)$ and $u_\m(z)$ as
\bea
   b(z) &=& b(x) + u^\m(x) k_b(r) \,\pa_\m b(x), \\
   u_\m(z) &=& u_\m(x) + u^\n(x) k_b(r) \,\pa_\n u_\m(x). \nono
\eea
The tensor $g_{\m\n}(z,\rt)$ we expand both in transverse coordinates and in $\rt$
\be
   g_{\m\n}(z,\rt) = g_{\m\n}(x,r) + u^\l(x) k_b(r) \,\pa_\l g_{\m\n}(x,r) + \rt_1(x,r) \,\pa_r g_{\m\n}(x,r). \nono
\ee
As mentioned above, the derivatives $\pa_\m b(x)$ can be converted into derivatives of the fluid velocity $\pa_\m u_\n(x)$ by the continuity equation
\be
   \pa_\m b = b \left( - \frac{1}{2\s-1} u_\m (\pa \cdot u) + (u \cdot \pa) u_\m \right),
\ee
which follows from the conservation of the perfect fluid energy-momentum tensor \eqref{pfTij}, $\pa^\m T_{0\m\n} =0$, or equivalently from the divergence equation \eqref{conf-div-eq} at order $\s$.

Putting everything together we obtain the transformation to first order in derivatives,
\bea
\label{1ordertrafo}
   \rt(x,r) &=& \rt_0(1+ k_b \frac{\pa \cdot u}{2\s-1}), \\
   z^\m(x,r) &=& x^\m + u^\m k_b +  u^\m \frac{\pa \cdot u}{2\s-1} l(r) + (u \cdot \pa) u^\m m(r), \nono
\eea
where $l(r)$, $m(r)$ satisfy the differential equations
\bea
   \frac{dl}{dr} &=&  -\frac{k_b(r)}{r^2 f_b(r)} - \frac{1}{r^3(f_b(r))^{3/2}}, \\
   \frac{dm}{dr} &=& -\frac{k_b(r)}{r^2f_b(r)} + \frac{1}{r^3 \sqrt{f_b(r)}}, \nono
\eea
with the boundary condition that they vanish for $r \rightarrow \infty$.
The metric in Eddington-Finkelstein coordinates up to first derivative order is then given by
\bea
\label{confEF1order}
   G_{\m\n}(x,r) &=& r^2[\eta_{\m\n} + (f_b(r)-1)u_\m u_\n] \\
     && - \chi_b(r) \s_{\m\n} - \frac{2 r}{2\s-1} u_\m u_\n (\pa \cdot u) + r (u \cdot \pa)(u_\m u_\n), \nono
\eea
where
\be
\label{exprchi}
   \chi_b(r) = \frac{2A(\rt_0(r))k_b(r) + \l(\rt(r))}{\rt_0(r)} = r^2 \left[2k_b(r) + \frac{2\g b}{\s} \log \frac{2-A^{\s/2}}{A^{\s/2}}\right].
\ee
Near the horizon $r\rightarrow 1/b$, the hypergeometric function in the definition of $k_b(r)$ in \eqref{hyper} develops a logarithmic divergence of the form \cite{AnS}
\be
   {}_2F_1(x,y;x+y;w) = -\frac{\G(x+y)}{\G(x)\G(y)} \log (w-1) + finite,
\ee
and $\chi_b(r)$ becomes
\be
   \chi_b \rightarrow \frac{(\g-1)}{\s b} \log (r-\frac{1}{b}) + finite.
\ee
Hence the divergence in $\chi_b(r)$ cancels precisely if $\g=1$, ie. the shear viscosity to entropy density bound $\eta_{2\s}/s_{2\s} \geq 1/4\pi$ is saturated.

In the non-conformal case, the transformation from Fefferman-Graham to Eddington-Finkelstein coordinates is given by the same coordinate transformations \eqref{0ordertrafo} and \eqref{1ordertrafo}. By either transforming \eqref{fullFG1ordernonconf} or by dimensionally reducing \eqref{confEF1order} according to the reduction ansatz \eqref{redans} we obtain the metric and scalar to first derivative order in Eddington-Finkelstein coordinates:
\bea
   G_{ij} &=& r^2[\eta_{ij} + (f_b(r) -1)u_i u_j] - \chi_b(r) \left[\s_{ij} + \frac{2\s-d}{(d-1)(2\s-1)}P_{ij} (\pa \cdot u)\right] \\
   && \qquad -\frac{2 r}{2\s-1} u_i u_j (\pa \cdot u) + r(u\cdot \pa) (u_i u_j), \nono \\
   \phi &=& \frac{2\s-d}{2} \left[ \log r^2 - \frac{1}{2\s-1} \frac{\chi_b(r)}{r^2} (\pa \cdot u) \right].
\eea
In particular, we see that the condition for the scalar and metric of the non-conformal solution to be smooth at the horizon is identical to the smoothness condition in the conformal case, namely that $\g=1$ in the definition of $\chi_b(r)$ in \eqref{exprchi}. Hence as expected also in the non-conformal case the bound $\eta_d/s_d \geq 1/4\pi$ is saturated for arbitrary $\s$.

In contrast, the fixed value of the ratio of the bulk to shear viscosity $\zeta_d/\eta_d = 2(1/(d-1) - c_s^2)$
does not follow from a smoothness condition but instead it follows from the equation of motions away from the horizon of the black brane. As mentioned in section \ref{univhydro} and in the introduction, it is a consequence of the generalized conformal structure established by the Ward identity \eqref{Wardid}.

\section{Discussion}

In this note we have shown that the universal sector of solutions asymptotic to non-conformal brane geometries can be obtained by dimensionally reducing asymptotically AdS pure gravity solutions and continuing the number of dimensions of the higher-dimensional geometry. As a consequence, the hydrodynamics dual to non-conformal black branes is fully determined in terms of the hydrodynamics dual to conformal black branes. We used this relation to rederive the first order contribution to the non-conformal hydrodynamic stress tensor and predict the second order contributions. As expected, the KSS bound \cite{KSSbound} for the ratio between shear viscosity and entropy density is always saturated. Furthermore we reconfirm that also the bound between shear and bulk viscosity proposed by \cite{Buchel:2007mf} is saturated. We show however that the saturation of this bound for non-conformal brane geometries follows from the generalized conformal symmetry, which indicates that it is of kinematical origin, unlike the KSS bound.

It would be interesting to explore whether the relation between conformal and non-conformal brane backgrounds also holds at the higher derivative level. Corrections to the KSS ratio for higher derivative bulk actions dual to conformal fluids have been investigated in \cite{higherd_etas}. The generalized conformal structure in non-conformal brane geometries is expected to hold for arbitrary coupling, although it is not clear whether
it would always descend from a higher dimensional conformal structure. For the cases of D4 branes
and fundamental strings, this would be the case since this is just the M-theory uplift, so at least in these cases the ratio of bulk to shear viscosity should not receive any corrections at the higher derivative level.

On a more general level, one might wonder which further generalization of the AdS/CFT dictionary can be found by compactifying asymptotically AdS spaces on other manifolds or more general tori. In such a setup, the lower-dimensional geometry will automatically inherit many holographic results from
the higher-dimensional asymptotically AdS case. Applied to hydrodynamics, one might obtain in this way non-conformal fluids with interesting properties.

Finally, it would be interesting to explore the hydrodynamics of non-conformal non-relativistic field theories. Hydrodynamics of conformal non-relativistic field theories have been explored by \cite{Rangamani:2008gi}. By compactification one might be able to obtain non-conformal generalizations.

\section*{Acknowledgements}

We would like to thank Paul McFadden and Marika Taylor for useful discussions. IK is supported by NWO, via the Vidi grant ``Holography, duality and time dependence in string theory''. KS would like to thank KITP
for hospitality during the completion of this work. This work was supported in part by the National Science
Foundation under Grant No. NSF PHY05-51164.

\appendix

\section{The asymptotic expansion of metric and scalar beyond the non-local mode} \label{appA}
\label{expnonlocal}

A general result that is most easily seen using the radial Hamiltonian formalism
\cite{Papadimitriou:2004ap, Papadimitriou:2004rz} is that a bulk solution is
uniquely specified by the holographic vevs. The reason is that the vevs are the radial
canonical momenta and the sources the corresponding coordinates. Thus specifying
the source and the vev is equivalent to specifying a point in the phase space of the theory,
which is equivalent to specifying a full solution.

A special case that is relevant for us is the gravity/scalar system for the non-conformal branes
with the boundary metric taken to be flat, $g\sub{0}_{ij} = \eta_{ij}$, and $\k\sub{0}=0$.
 Then all subleading terms in the expansion \eqref{metrscalexp} up to order $\s$ including the logarithmic term vanish, since they depend on derivatives of the sources $g\sub{0}$ and $\k\sub{0}$. The order $\s$ terms will be given by
\be
   g\sub{2\s}_{ij} = \frac{1}{2\s L} T_{ij}, \qquad \k\sub{2\s} = -\frac{1}{4\s L}  T^i_i,
\ee
where $T_{ij}$ and $T^i_i$ denote the vev of the dual energy-momentum tensor and its trace, and $\k\sub{2\s}$ is determined by the requirement that the generalized conformal Ward identity \eqref{Wardid} is satisfied.
The higher order terms in Fefferman-Graham expansion are then determined in terms of $T_{ij}$.
We will need the schematic form of these coefficients.
The non-linear equations of motion induce the expansion
\bea
   g(z,\rho)_{ij} &=& \eta_{ij} + \sum_{\tau = n\s + m} \r^\tau g\sub{2\tau}_{ij}, \\
   \k(z,\rho) &=& \k\sub{0} + \sum_{\tau = n\s +m} \r^\tau \k\sub{2\tau}, \nono
\eea
where $n>1$ and $m>0$ in the summation are positive integers. Suppressing the index structure, the higher order terms are schematically of the form
\be
\label{gsub2tau}
   g\sub{2\tau}_{ij} \propto \sum_{n\s+m = \tau} a_{n,m}(\pa^{2m} T^n)_{ij},
\ee
and similarly for $\k\sub{2n\s+2m}$, due to dimensional considerations. In particular, if $T_{ij}$ is constant, only the coefficients with $\tau = n\s$ are non-zero. Once $T_{ij}$ becomes dependent on the boundary coordinates, the transverse derivatives in \eqref{gsub2tau} always enter in pairs. This fact is used in section \ref{GenbbFG} to restrict the form of the first order derivative correction to the metric.

\end{document}